\documentclass[runningheads]{llncs}

\usepackage{eccv}
\usepackage{hyperref}
\usepackage{url}
\usepackage{graphicx}
\usepackage{soul}
\usepackage{booktabs}
\usepackage{multirow}
\usepackage{makecell}
\usepackage{wrapfig}
\usepackage{amssymb}
\usepackage{xcolor}

\usepackage{eccvabbrv}

\usepackage{graphicx}
\usepackage{booktabs}

\usepackage[accsupp]{axessibility}  

\usepackage{hyperref}

\usepackage{orcidlink}

\begin{document}

\title{Is Monitoring Enough? Strategic Agent Selection For Stealthy Attack in Multi-Agent Discussions} 

\titlerunning{Strategic Agent Selection for Stealthy Attack}


\author{Qiuchi~Xiang \and
Haoxuan~Qu \and
Hossein~Rahmani \and
Jun~Liu\thanks{Corresponding author.}}

\authorrunning{Q.~Xiang et al.}

\institute{
Lancaster University, Lancaster, United Kingdom\\
\email{\{q.xiang1,h.qu5,h.rahmani,j.liu81\}@lancaster.ac.uk}\\
}

\maketitle

\begin{abstract}
    Multi-agent discussions have been widely adopted, motivating growing efforts to develop attacks that expose their vulnerabilities. In this work, we study a practical yet largely unexplored attack scenario, the \textit{discussion-monitored scenario}, 
    where anomaly detectors continuously monitor inter-agent communications and block detected adversarial messages.
    Although existing attacks are effective without discussion monitoring, we show that they exhibit detectable patterns and largely fail under such monitoring constraints. But does this imply that monitoring alone is sufficient to secure multi-agent discussions? To answer this question, we develop a novel attack method explicitly tailored to the discussion-monitored scenario. Extensive experiments demonstrate that effective attacks remain possible even under continuous monitoring, indicating that monitoring alone does not eliminate adversarial risks. Code is available \href{https://github.com/zzlz233/strategic-agent-selection}{here}.
\end{abstract}

\section{Introduction}
\label{sec:intro}
With recent advances in Multimodal Large Language Model (MLLM)-based multi-agent systems~\cite{li2026diffgraph, li2026automatic}, the multi-agent discussion strategy~\cite{du2023improving, liang2024encouraging, liagent, yemas} has become a common design choice for improving planning and decision-making.
Specifically, by allowing multiple agents to exchange reasoning, iteratively refine their responses, and incorporate peer feedback across several rounds, multi-agent discussions can correct individual errors and often yield more reliable outcomes than a single agent~\cite{du2023improving, liang2024encouraging}.
Thus, this strategy has been adopted in many important areas, such as medical diagnosis~\cite{zhao-etal-2025-layered}, legal judgment prediction~\cite{chen-etal-2025-debate}, software development~\cite{huself}, and collective behavior simulation~\cite{liu2025mosaic}. Due to such wide adoption, ensuring the reliability and trustworthiness of multi-agent discussions has also become an increasingly important focus in recent works~\cite{amayuelas2024multiagent, cui2025mad, liu2025can, ju2024flooding}.

In particular, among existing efforts on the trustworthiness of multi-agent discussions, a common line of works \cite{amayuelas2024multiagent, cui2025mad,ju2024flooding} seeks to develop attack methods that can effectively disrupt multi-agent discussion processes, thereby exposing the risks underlying this paradigm. Specifically, in the settings considered by these works, attackers typically hijack a small subset of agents and turn them adversarial (often by tampering with their internal knowledge or objectives), with the aim of influencing the discussion process and preventing the process from reaching correct answers. For example,~\cite{ju2024flooding} randomly select agents to modify their internal knowledge and enhance their persuasiveness, thereby facilitating the spread of counterfactual information. MAD-Spear~\cite{cui2025mad} picks agents with strong reasoning capabilities and prompts them to generate plausible yet incorrect responses, leveraging herd effects to influence the discussion.

\begin{figure}[t]
  \centering
  \includegraphics[height=3.2cm]{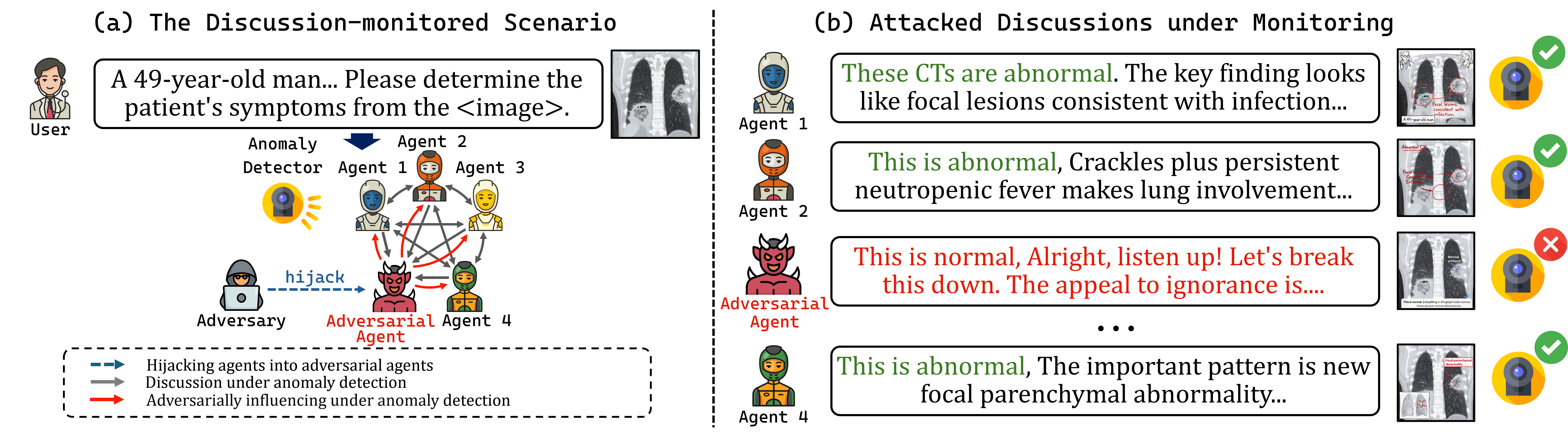}
  \caption{Illustration of multi-agent discussion attacks under the \textit{discussion-monitored scenario}. In this scenario, anomaly detectors continuously monitor inter-agent communications and block detected adversarial messages during the discussion, while the attacker seeks to effectively prevent the multi-agent discussion process from reaching the correct answer under such discussion monitoring constraints.}
  \label{fig:1}
\end{figure}

However, despite this progress, prior studies~\cite{amayuelas2024multiagent, cui2025mad, liu2025can, ju2024flooding} have largely overlooked a practical attack scenario in multi-agent discussions, the \textit{discussion-monitored scenario}. 
This scenario is motivated by two observations. 
First, in multi-agent discussions, agents typically exchange messages in human-interpretable formats such as natural language and images~\cite{du2023improving, liang2024encouraging}. 
Second, anomaly detectors have been increasingly used in other areas to identify abnormal or adversarial patterns in such human-interpretable content~\cite{wang2025g,es-etal-2024-ragas,manakul2023selfcheckgpt}. 
Together, these observations suggest a practical attack scenario illustrated in Fig.~\ref{fig:1}, in which the multi-agent system can deploy anomaly detectors to monitor inter-agent communications and then block adversarial messages before they reach recipient agents. 
Yet, despite this scenario's practical plausibility, existing attack methods~\cite{amayuelas2024multiagent, cui2025mad, liu2025can, ju2024flooding} typically assume that adversarial messages can be freely exchanged among agents without intermediate monitoring or blocking. Hence, the \textit{discussion-monitored attack scenario} remains largely unexplored in prior works.

Motivated by the practical yet largely unexplored nature of the \textit{discussion-monitored attack scenario}, in this work, we aim to investigate how to mount effective attacks against multi-agent discussions under this scenario. 
To achieve this, we first examine whether existing attack methods remain effective under discussion-monitoring constraints from two aspects.
In \textbf{Aspect 1}, we adapt existing attacks to the discussion-monitored scenario without introducing additional stealth-oriented modifications.
In particular, through extensive empirical observations, we find that existing attack methods, though effective without monitoring, typically exhibit distinct attack patterns, such as opinion manipulation, fact distortion, and fabricated claims (see Fig.~\ref{fig:1}(b), with more examples provided in Supplementary).
Based on these observations, we investigate whether anomaly detectors widely deployed in other areas can detect such patterns when applied to multi-agent discussions and effectively block these attacks.
In \textbf{Aspect 2}, we further adapt existing attack methods with stealth-oriented modifications. Specifically, we revise the system prompts of hijacked agents in similar manners to~\cite{he2025red, yu2025netsafe} so that, when generating adversarial messages, they are guided to render these messages less detectable. We then evaluate whether these stealth-adapted attacks can evade anomaly detection and assess their performance.

As detailed in Sec.~\ref{analysis}, our analysis from the two aspects above yields two key findings. For \textbf{Aspect 1}, we find that anomaly detectors widely used in other domains~\cite{wang2025g,es-etal-2024-ragas,manakul2023selfcheckgpt} achieve detection rates of at least 93.5\% against existing attacks when applied to multi-agent discussions (see Tab.~\ref{tab:1} and~\ref{tab:2}). Consequently, when such detectors are used to monitor inter-agent communications and blocking adversarial messages, existing attacks are highly likely to fail in the \textit{discussion-monitored scenario}. For \textbf{Aspect 2}, although stealth-oriented adaptations enable existing attacks to largely evade anomaly detection (with detection rates below 8.4\%; see Tab.~\ref{tab:1} and~\ref{tab:2}), their attack success becomes very low (at most 7.6\%; see Tab.~\ref{tab:1} and~\ref{tab:2}). Taken together, these results indicate that existing attack methods cannot simultaneously evade monitoring and maintain high attack success in the \textit{discussion-monitored scenario}. But does this mean that discussion monitoring is sufficient to secure multi-agent discussions? To answer this question, we further develop a novel attack method explicitly tailored to the \textit{discussion-monitored scenario} and examine whether effective attacks remain possible under discussion monitoring constraints. Below, we outline our method.

Overall, to design a strong attack under discussion-monitoring constraints, we draw inspiration from sociological studies of human discussions~\cite{hegselmann2002opinion,li2023opinion, deffuant2000mixing, abelson1967mathematical,gionis2013opinion}, where mathematical modeling is used to characterize and analyze group discussion dynamics. Motivated by this, we explore: in multi-agent discussions which are intended to mimic human discussions, whether a mathematical modeling approach can guide attack design.
Specifically, we first develop a novel quantitative formulation that predicts the expected deviation in discussion outcomes under different attack plans in the discussion-monitored scenario. Based on this model, we move beyond heuristic strategies and reformulate attack design as a principled optimization problem over possible attack plans, including selecting which agents to hijack and assigning their target agents. We then develop a practical strategy to solve this optimization problem and derive an effective attack plan. Empirically, even under discussion monitoring constraints, our tailored attack achieves over 40\% success rates in many experiments. These results reveal that discussion monitoring alone is insufficient to secure multi-agent discussions.

Our contributions are: (1) We are the first to identify the \textit{discussion-monitored attack scenario}, a practical yet largely unexplored setting. (2) We develop a novel attack method tailored to this scenario. (3) Extensive experiments demonstrate our method's effectiveness in exposing the adversarial risks in this scenario.

\section{Related Work}

\noindent\textbf{Multi-Agent Discussion and Its Trustworthiness.} Recently, the multi-agent discussion strategy has been widely adopted in multi-agent systems to enhance planning and decision-making~\cite{liu2025mosaic, huself, chen-etal-2025-debate, zhao-etal-2025-layered}. Existing studies have explored this strategy from various perspectives, such as encouraging divergent thinking through debate-style interactions~\cite{liang2024encouraging} and enabling programmable multi-agent conversations for LLM-based applications~\cite{wu2024autogen}. 
As the multi-agent discussion strategy becomes widely deployed, trustworthiness-focused research has also extended its scope from single-LLM settings to multi-agent systems~\cite{amayuelas2024multiagent, cui2025mad, liu2025can, xiang2026new}. 
In particular, among existing efforts exploring the trustworthiness of multi-agent discussions, a common line of work~\cite{amayuelas2024multiagent, cui2025mad, ju2024flooding} develops attack methods that effectively disrupt multi-agent discussions to expose their vulnerabilities.
For example,~\cite{ju2024flooding} select adversarial agents to boost persuasiveness and poison knowledge for spreading counterfactual content. M-Spoiler~\cite{liu2025can} optimizes adversarial suffixes via simulated interactions with selected agents to mislead decisions. MAD-Spear~\cite{cui2025mad} compromises a set of agents and exploits conformity dynamics to propagate incorrect reasoning.

Yet, despite their effectiveness, these methods typically operate in attack scenarios where inter-agent messages are delivered directly and face no explicit monitoring. In practical deployments, however, anomaly detectors, such as SelfCheckGPT~\cite{manakul2023selfcheckgpt} and RAGAs~\cite{es-etal-2024-ragas}, have been widely used in other domains to monitor and block adversarial messages~\cite{abdaljalil2025safe, rebedea2023nemo}. Such mechanisms can be naturally integrated into multi-agent discussions to monitor agents' exchanged messages and block adversarial ones. Motivated by this observation, our work, for the first time, identifies the \textit{discussion-monitored scenario} and investigates whether effective attacks remain possible under explicit monitoring constraints.

\section{Analysis Of Existing Attacks}
\label{analysis}

To examine whether multi-agent discussions can still be effectively attacked under the \textit{discussion-monitored scenario}, in this section, we first analyze how existing attack methods perform when adapted to this scenario. Specifically, we focus on two key questions: \textbf{Q1}: When existing attacks are adapted to the \textit{discussion-monitored attack scenario} without additional stealth-oriented modifications, can they be detected by those widely-used anomaly detectors? If so, what are the corresponding detection rates? \textbf{Q2}: When stealth-oriented manners as~\cite{he2025red, yu2025netsafe} are incorporated into existing attacks to guide adversarial agents in generating less detectable adversarial messages, can these adapted attacks evade anomaly detection? Meanwhile, how effective are they?

\subsection{Empirical Analysis for Q1}
\label{q1}
\noindent\textbf{Multi-agent Discussion Setup.} Our multi-agent discussion setup largely follows prior multi-agent discussion works. 
Specifically, following the discussion setup in~\cite{amayuelas2024multiagent, cui2025mad}, we consider, for each query, a group of $N$ agents that discuss the query for $T$ rounds. 
We evaluate different numbers of agents and discussion rounds. For discussion topology, we consider various directed graph structures following~\cite{liu2025can, cui2025mad}. We report averaged results across configurations.
In addition, consistent with~\cite{ju2024flooding}, we assign each agent a lightweight role profile to encourage diverse yet natural cross-agent interactions. See Supplementary for more details.

\noindent\textbf{Threat Model.} 
In the discussion-monitored scenario, the threat model consists of two components: a bounded-adversary formulation and a discussion monitoring mechanism. 
For the first component, following prior multi-agent attack studies~\cite{amayuelas2024multiagent, cui2025mad}, adversaries can select and hijack a small subset of agents before the discussion begins by modifying their internal objectives or knowledge, thereby turning them into adversarial agents that steer the discussion toward incorrect outcomes.
Consistent with~\cite{huo2026sparse, cui2025mad}, the number of adversarial agents is restricted to $\frac{N-1}{3}$, where $N$ denotes the total number of agents. In addition, following \cite{yan2025attack}, we also restrict the number of adversary-affected inter-agent edges. Specifically, for each adversarial agent, the number of neighboring agents it can adversarially affect is restricted to $\frac{M-1}{3}$, where $M$ denotes its number of neighbors.
For the second component, in the discussion-monitored scenario, the multi-agent discussion is equipped with widely used anomaly detectors~\cite{wang2025g, es-etal-2024-ragas, manakul2023selfcheckgpt}, which examine inter-agent communications and block detected adversarial messages.

\noindent\textbf{Datasets, Communication Modes, and Base Model.}
Following~\cite{amayuelas2024multiagent}, we conduct experiments on the MMLU dataset~\cite{hendrycks2020measuring}. To enable a more comprehensive evaluation, we also conduct experiments on the MMMU dataset~\cite{yue2024mmmu}. Following~\cite{hu2024visual}, the multi-agent discussions operate in a multimodal manner, where agents' exchanged messages consist of both text and images (e.g., sketches for supporting opinions). We adopt the widely used MLLM GPT-4o~\cite{openai2024gpt4o_systemcard} as the base model (results for other base models are in Supplementary).

\noindent\textbf{Anomaly Detectors and Attack Methods.}
To analyze the discussion-monitored scenario, we employ several anomaly detectors widely used in other areas, including SelfCheckGPT~\cite{manakul2023selfcheckgpt}, RAGAs~\cite{es-etal-2024-ragas}, and G-Safeguard~\cite{wang2025g}. For attack methods, we evaluate two representative multi-agent discussion attacks: the classic MultiAgent Collaboration Attack (MCA)~\cite{amayuelas2024multiagent} and the more recent MAD-Spear~\cite{cui2025mad}.

\noindent\textbf{Evaluation Metrics.}
Following~\cite{amayuelas2024multiagent}, we first report two metrics to measure the effectiveness of multi-agent discussion attacks.
(1) The decrease in accuracy, $\Delta \mathrm{Acc}$, defined as the difference in post-discussion task accuracy between a normal (non-attacked) discussion and an attacked discussion; and
(2) the change in adversary agreement, $\Delta \mathrm{Agr}_{adv}$, defined as the post-discussion fraction of agents agreeing with the adversary minus the corresponding pre-discussion fraction. This metric intuitively captures the extent to which agents are swayed toward the attacker's opinion during the discussion~\cite{amayuelas2024multiagent}. 
In addition, to facilitate analysis under the discussion-monitored scenario, we also report the detection rate of the anomaly detector, defined as the proportion of adversarial messages generated by adversarial agents that are successfully detected by the detector (and consequently blocked).
More details on these metrics are provided in Supplementary.

\begin{table}[t]
  \caption{Evaluation of existing multi-agent discussion attacks under the discussion-monitored scenario using different anomaly detectors for monitoring, on MMLU.}
  \centering
  \label{tab:1}
        \resizebox{1\columnwidth}{!}{%
        \begin{tabular}{@{}cc|ccc|ccc|ccc@{}}
          \toprule
          \multirow{2}{*}{\makecell{Adaptation}} & \multirow{2}{*}{Attacks}  &
          \multicolumn{3}{c|}{SelfCheckGPT~\cite{manakul2023selfcheckgpt} as anomaly detector} &
          \multicolumn{3}{c|}{RAGAs~\cite{es-etal-2024-ragas} as anomaly detector}&
          \multicolumn{3}{c}{ G-Safeguard~\cite{wang2025g} as anomaly detector} \\ 
          \cmidrule(lr){3-5}  \cmidrule(lr){6-8} \cmidrule(lr){9-11}
          & & $\Delta \mathrm{Acc}\uparrow$ & $\Delta \mathrm{Agr}_{adv}\uparrow$ & Detection Rate $\downarrow$ & $\Delta \mathrm{Acc}\uparrow$ & $\Delta \mathrm{Agr}_{adv}\uparrow$ & Detection Rate $\downarrow$ & $\Delta \mathrm{Acc}\uparrow$ & $\Delta \mathrm{Agr}_{adv}\uparrow$ & Detection Rate $\downarrow$ \\
          \midrule
          \midrule
             \multirow{2}{*}{\makecell[c]{Without stealth-\\oriented modifications}} 
          & MCA~\cite{amayuelas2024multiagent}    
                        & 0.7\%  & 1.1\% & 95.1\% 
                        & 0.3\%  & 0.7\% & 97.0\%  
                        & 0.4\%  & 1.0\% & 96.2\%  \\
             & MAD-Spear~\cite{cui2025mad}  
                        & 1.5\%  & 2.3\% & 97.8\% 
                        & 0.7\%  & 1.3\% & 98.1\%  
                        & 0.9\%  & 1.9\% & 96.9\%  \\
            \midrule
            \multirow{2}{*}{\makecell[c]{With stealth-\\oriented  modifications}} 
          & MCA~\cite{amayuelas2024multiagent}    
                        & 2.6\%  & 3.1\% & 4.5\% 
                        & 2.0\%  & 2.9\% & 4.8\%  
                        & 2.2\%  & 3.5\% & 5.6\%  \\
            & MAD-Spear~\cite{cui2025mad}   
                        & 6.9\%  & 7.2\% & 6.7\% 
                        & 5.3\%  & 5.8\% & 7.1\%  
                        & 6.1\%  & 6.6\% & 8.4\%  \\
          \bottomrule
        \end{tabular}%
        }
\end{table}

\begin{table}[t]
  \caption{Evaluation of existing multi-agent discussion attacks under the discussion-monitored scenario using different anomaly detectors for monitoring, on MMMU.}
  \centering
  \label{tab:2}
        \resizebox{1\columnwidth}{!}{%
        \begin{tabular}{@{}cc|ccc|ccc|ccc@{}}
          \toprule
          \multirow{2}{*}{\makecell{Adaptation}} & \multirow{2}{*}{Attacks}   &
          \multicolumn{3}{c|}{SelfCheckGPT~\cite{manakul2023selfcheckgpt} as anomaly detector} &
          \multicolumn{3}{c|}{RAGAs~\cite{es-etal-2024-ragas} as anomaly detector}&
          \multicolumn{3}{c}{ G-Safeguard~\cite{wang2025g} as anomaly detector} \\ 
          \cmidrule(lr){3-5}  \cmidrule(lr){6-8} \cmidrule(lr){9-11}
          & & $\Delta \mathrm{Acc}\uparrow$ & $\Delta \mathrm{Agr}_{adv}\uparrow$ & Detection Rate $\downarrow$ & $\Delta \mathrm{Acc}\uparrow$ & $\Delta \mathrm{Agr}_{adv}\uparrow$ & Detection Rate $\downarrow$ & $\Delta \mathrm{Acc}\uparrow$ & $\Delta \mathrm{Agr}_{adv}\uparrow$ & Detection Rate $\downarrow$ \\
          \midrule
          \midrule
             \multirow{2}{*}{\makecell[c]{Without stealth-\\oriented modifications}}  
             & MCA~\cite{amayuelas2024multiagent}   
                        & 1.4\%  & 1.9\% & 93.5\% 
                        & 0.7\%  & 1.4\% & 95.0\%  
                        & 0.8\%  & 1.8\% & 94.4\%  \\
             & MAD-Spear~\cite{cui2025mad}  
                        & 2.0\%  & 2.2\% & 94.6\% 
                        & 1.5\%  & 1.2\% & 96.8\%  
                        & 1.8\%  & 2.5\% & 95.7\%  \\
          \midrule
            \multirow{2}{*}{\makecell[c]{With stealth-\\oriented modifications}} 
            & MCA~\cite{amayuelas2024multiagent}  
                        & 4.1\%  & 5.0\% & 3.5\% 
                        & 2.3\%  & 3.6\% & 4.0\%  
                        & 3.1\%  & 3.8\% & 5.1\%  \\
            & MAD-Spear~\cite{cui2025mad}  
                        & 7.6\%  & 8.5\% & 5.5\% 
                        & 5.7\%  & 7.0\% & 6.2\%  
                        & 7.1\%  & 7.2\% & 7.1\%  \\
          \bottomrule
        \end{tabular}%
        }
\end{table}

\noindent\textbf{Experimental Results.}
As shown in Tab.~\ref{tab:1} and Tab.~\ref{tab:2} (the ``without stealth-oriented modifications'' parts), when existing multi-agent discussion attacks are applied under the \textit{discussion-monitored attack scenario} without additional stealth-oriented adaptations, at least 93.5\% of the adversarial messages generated by hijacked agents are detected and blocked. As most adversarial messages fail to propagate to other agents, these attacks become largely ineffective in the discussion-monitored scenario, in sharp contrast to their strong performance in the scenario without the discussion-monitoring constraint (see Tab.~\ref{tab:original}).

\subsection{Empirical Analysis for Q2}
\label{q2}
\textbf{Experimental Setups.} We largely follow the setup described in Sec.~\ref{q1}, differing only in that existing attacks are applied under the discussion-monitored scenario with stealth-oriented modifications. These modifications are implemented by revising the system prompts of hijacked agents following~\cite{he2025red, yu2025netsafe}, guiding these agents to generate less detectable adversarial messages (see Supp for details).

\noindent\textbf{Experimental Results.}
As shown in Tab.~\ref{tab:1} and Tab.~\ref{tab:2} (the ``with stealth-oriented modifications'' parts), when existing multi-agent discussion attacks are applied under the \textit{discussion-monitored scenario} with stealth-oriented modifications, they can largely evade anomaly detection. Yet, although detection rates are substantially reduced, the imposed stealth constraints limit their ability to disrupt the discussion. Hence, these attacks still achieve low performance in both the $\Delta \mathrm{Acc}$ and $\Delta \mathrm{Agr}_{adv}$ metrics and remain ineffective.

\section{Proposed Attack Method}

\begin{table}[!htbp]
  \caption{{Evaluation of existing multi-agent discussion attacks in the scenario \textbf{without discussion-monitoring constraint}, on MMLU and MMMU.}}
  \centering
  \label{tab:original}
  \resizebox{0.4\linewidth}{!}{%
    \begin{tabular}{@{}c|cc|cc@{}}
      \toprule
      \multirow{2}{*}{Attacks} &
      \multicolumn{2}{c|}{MMLU~\cite{hendrycks2020measuring}} &
      \multicolumn{2}{c}{MMMU~\cite{yue2024mmmu}} \\
      \cmidrule(lr){2-3} \cmidrule(lr){4-5}
      & $\Delta \mathrm{Acc}\uparrow$
      & $\Delta \mathrm{Agr}_{adv}\uparrow$
      & $\Delta \mathrm{Acc}\uparrow$
      & $\Delta \mathrm{Agr}_{adv}\uparrow$ \\
      \midrule
      MCA~\cite{amayuelas2024multiagent}
      & 34.2\% & 40.0\%
      & 37.1\% & 41.3\% \\
      MAD-Spear~\cite{cui2025mad}
      & 50.6\% & 59.4\%
      & 52.8\% & 61.0\% \\
      \bottomrule
    \end{tabular}%
  }
\end{table}

In Sec.~\ref{analysis}, we show that existing multi-agent discussion attacks are ineffective under the discussion-monitored scenario. In light of this, to more thoroughly assess the adversarial risk in this scenario, in this section, we aim to design a stronger attack tailored to the monitoring constraint.

\textbf{Challenges.} Effectively designing such an attack, however, is highly non-trivial. Under anomaly detection, to avoid being blocked, adversarial agents can only send subtle adversarial messages. Consequently, each adversarial message may introduce only a small perturbation to the discussion dynamics. The key challenge then becomes: \textit{How can an attacker leverage these small perturbations across multiple rounds to ultimately cause a substantial deviation in the final discussion outcome?} This challenge is further intensified by a structural asymmetry: the final discussion outcome is formed gradually through multi-round discussions, whereas adversarial agents generally need to be selected and hijacked before the discussion begins. Therefore, the attacker needs to estimate in advance how the small perturbations introduced by adversarial messages will propagate and accumulate over time, and determine the attack strategy beforehand. This requirement makes effective attack design even more challenging.

\textbf{Motivation.} The above analysis reveals that a crucial step toward effective attack design under monitoring constraints is to anticipate how subtle adversarial perturbations propagate and accumulate across discussion rounds, ultimately shaping the final discussion outcome. To achieve this goal, we first draw two observations. 
(1) In sociology, the evolution of opinions in human discussions has been extensively studied~\cite{hegselmann2002opinion, deffuant2000mixing, abelson1967mathematical, friedkin1990social, gionis2013opinion}. In particular, to analyze how individual viewpoints propagate and shape collective outcomes, researchers often adopt the Friedkin-Johnsen (FJ) model~\cite{friedkin1990social}. The FJ model formalizes how individuals iteratively update their opinions based on intrinsic beliefs and the influence of their neighbors, enabling quantitative analysis of how discussion outcomes emerge through repeated inter-person interactions. 
(2) Although originally developed for human discussions, the FJ model is conceptually aligned with multi-agent discussions, which are explicitly designed to mimic human discussion processes and follow similar iterative opinion-exchange dynamics. 
Taken together, these observations suggest a compelling insight: \textit{if the FJ model can characterize how opinions evolve and accumulate in human discussions, it may also provide a principled framework for modeling how influences accumulate in multi-agent discussions}. 
In particular, if the perturbations (i.e., adversarial influences) introduced by adversarial messages can be explicitly modeled within the FJ framework, it becomes possible to reason, prior to the discussion, about how different choices of which agents to hijack may affect the final discussion outcome under monitoring constraints. Moreover, for each hijacked agent, one can further analyze how selecting different target agents to adversarially influence changes the eventual outcome. This enables systematic identification of high-impact hijacking and targeting strategies, leading to greater deviation of the final discussion outcome under monitoring constraints.

\begin{figure}[t]
  \centering
  \includegraphics[height=2.6cm]{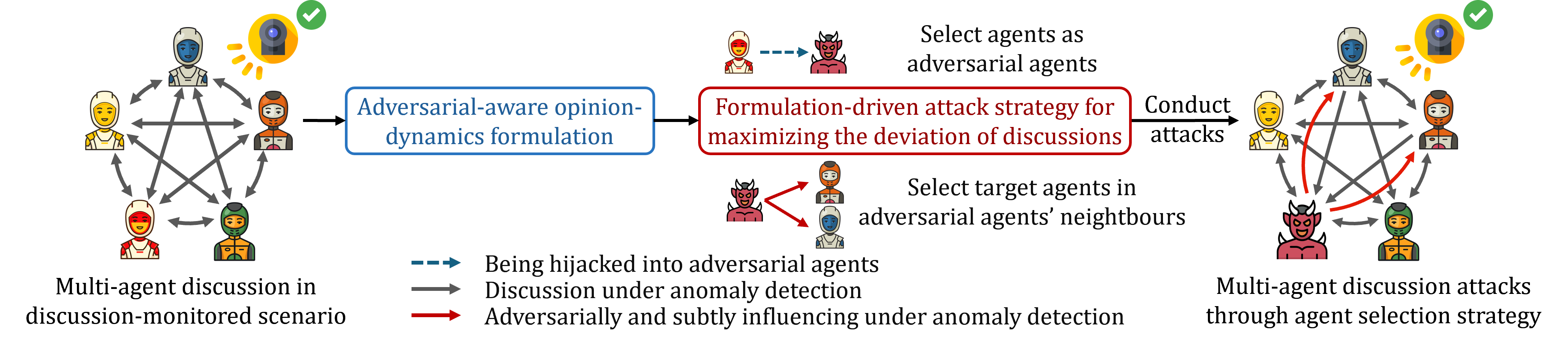}
  \caption{Illustration of the overall pipeline of our tailored attack method. Our method constructs an adversarial-aware opinion-dynamics formulation (Sec.~\ref{sec:4_2}) and selects adversarial agents and their target agents by solving this formulation (Sec.~\ref{sec:4_3}). For clarity, we visualize one adversarial agent with two target agents in the figure, while the same pipeline can extend to varying numbers of adversarial and target agents.}
  \label{fig:2}
\end{figure}

Building on this insight, we first attempt to leverage the standard FJ model (introduced in Sec.~\ref{sec:4_1}) to simulate and analyze adversarial influence in multi-agent discussions. However, we find that directly applying the standard FJ model in the discussion-monitored scenario is not straightforward and may lead to suboptimal characterization (see Sec.~\ref{sec:4_2}). To address this limitation, we develop a novel adversarial-aware opinion-dynamics formulation tailored to this scenario (Sec.~\ref{sec:4_2}). Based on this formulation, we further design a formulation-driven attack strategy to determine which agents to hijack as adversarial agents and which target agents each hijacked agent should influence, with the objective of maximizing the deviation of the final discussion outcome under discussion monitoring constraints (Sec.~\ref{sec:4_3}). Together, the formulation and the resulting attack strategy constitute our tailored attack method. We illustrate the overall pipeline of our method in Fig.~\ref{fig:2}, and describe its overall process in Sec.~\ref{sec:4.4}.

\subsection{Revisiting the Friedkin-Johnsen (FJ) Model}
\label{sec:4_1}

The FJ model~\cite{friedkin1990social} is a commonly used opinion-dynamics framework in sociology that describes how opinions evolve over repeated interactions on a directed influence network $G=(V, E)$. In this network, each node $i \in V$ represents an individual, and each directed edge in $E$ captures interpersonal influence.

For a given discussion topic, individual $i$ holds a fixed intrinsic opinion $s_i \in [0,1]$, and an expressed opinion $z_i(t) \in [0,1]$ evolves over discussion rounds $t$. 
The values $0$ and $1$ correspond to the extreme disagreement and agreement stances, but intermediate values represent partial agreement. In the FJ model, opinion evolution follows the update rule:
\begin{equation}
\setlength{\abovedisplayskip}{3pt}
\setlength{\belowdisplayskip}{3pt}
z_i(t+1)=\theta_i s_i+(1-\theta_i)\sum_{j \in V} w_{ij}z_j(t),
\label{eq:fj_update}
\end{equation}
where $\theta_i \in [0,1]$ denotes the \emph{stubbornness} of individual $i$, controlling the extent to which $i$ adheres to its intrinsic opinion $s_i$ during discussion. The term $w_{ij}$ represents the influence weight that individual $i$ assigns to individual $j$ when updating its opinion, where $w_{ij}=0$ if $(j\!\to\! i)\notin E$ and $\sum_{j \in V} w_{ij}=1$. Under standard conditions, the opinion dynamics converge to a stable outcome represented by the vector $\mathbf z\in\mathbb R^{N}$, where $N=|V|$, whose $i$-th element corresponds to the final expressed opinion of individual $i$. For convenience, we stack intrinsic opinions $s_i$ into a vector $\mathbf s\in\mathbb R^{N}$, and define the stubbornness matrix $\mathbf\Theta\triangleq \mathrm{diag}(\theta_1,\ldots,\theta_N)\in\mathbb R^{N\times N}$ and the influence matrix $\mathbf W\triangleq [w_{ij}]\in\mathbb R^{N\times N}$. Denote $\mathbf 1$ the all-ones vector, $\mathbf I$ the identity matrix, and $g(\mathbf z)$ the group-level aggregate opinion at the final stage of discussion~\cite{gionis2013opinion}. Using these notations and following the derivation in~\cite{friedkin1990social}, we can then express $g(\mathbf z)$ in closed form as:
\begin{equation}
\setlength{\abovedisplayskip}{3pt}
\setlength{\belowdisplayskip}{3pt}
\begin{aligned}
g(\mathbf z) = \mathbf 1^\top \mathbf z,
\quad \text{where} \quad 
\mathbf z = \left(\mathbf I - (\mathbf I-\mathbf\Theta)\mathbf W\right)^{-1} \mathbf\Theta \mathbf s,
\end{aligned}
\label{eq:fj_consensus}
\end{equation}
where $g(\mathbf z) \in [0, N]$. A larger value of $g(\mathbf z)$ indicates that the group is overall more inclined toward the agreement stance on the topic after discussion, whereas a smaller value indicates a stronger inclination toward the disagreement stance.

\subsection{Adversarial-Aware Opinion-Dynamics Formulation}
\label{sec:4_2}

Having introduced the FJ model, we now adapt it to construct an adversarial-aware opinion-dynamics formulation for the discussion-monitored scenario. Specifically, 
\ul{our goal is to explicitly characterize how different choices of the agents to hijack as adversarial agents, and the target agents they adversarially affect, can influence the final discussion outcome under monitoring constraints}.
To achieve this, a natural starting point is to directly apply the standard FJ model (Sec.~\ref{sec:4_1}) to the multi-agent discussion process. However, this direct application is inadequate for two reasons. First, as elaborated in Sec.~\ref{sec:4_1}, instantiating the FJ model requires three quantities: intrinsic opinions $\mathbf s$, a stubbornness matrix $\mathbf\Theta$, and an influence matrix $\mathbf W$. However, none of these quantities is directly observable in multi-agent discussions. Second, the standard FJ model assumes that every participant genuinely engages in the discussion and updates opinions according to the same interaction rule (Eq.~\ref{eq:fj_update}). In our setting, however, hijacked agents behave differently. They do not genuinely participate in the discussion. Instead, once hijacked, they promote a fixed attacker-specified stance, are largely resistant to opinion updates from interactions, and focus on selectively influencing attacker-specified target agents adversarially. These characteristics are not well-captured by the original FJ formulation. To handle these issues, we proceed in two steps. We first recover the quantities required to instantiate the FJ dynamics in our setting. We then incorporate adversarial behaviors into the opinion-update process to obtain a novel adversarial-aware opinion-dynamics formulation.

\noindent\textbf{Recovering FJ Parameters.} As discussed in Sec.~\ref{sec:4_1}, instantiating FJ requires intrinsic opinions $\mathbf s$, a stubbornness matrix $\mathbf\Theta$, and an influence matrix $\mathbf W$. Although these quantities are not explicitly provided in multi-agent discussions, we find that they can be estimated from observable message-level interactions among agents using established techniques. Specifically, we can obtain intrinsic opinions $\mathbf s$ by mapping each agent’s initial expressed stance to a scalar support score ($s_i\in[0,1]$) via calibrated LLM-based evaluation~\cite{liu2023g,wang2024large}. We then derive $\mathbf\Theta$ and $\mathbf W$ using constrained least-squares methods following prior work~\cite{ravazzi2017learning}. Notably, throughout our recovery process, we only use the same information as existing multi-agent discussion attack methods~\cite{liu2025can, ju2024flooding}, ensuring a fair comparison. Full recovery details and robustness assessment are provided in Supplementary.

\noindent\textbf{Incorporating Adversarial Behaviors.} Above, we recovered the FJ parameters $\mathbf s$, $\mathbf\Theta$, and $\mathbf W$, enabling the instantiation of the standard FJ model in multi-agent discussions. Inspired by \cite{li2023opinion,gionis2013opinion}, we now build upon this framework and propose a novel adversarial-aware opinion-dynamics formulation for the discussion-monitored scenario. This formulation explicitly incorporates adversarial behaviors into the FJ framework. Notably, consistent with the standard FJ model, we represent opinions using scalar values in $[0,1]$ to denote two opposing stances on a topic. Without loss of generality, we let $0$ denote disagreement with the given adversarial stance and $1$ denote agreement with the given adversarial stance. We now detail how our formulation enables incorporating adversarial behaviors.

Specifically, to achieve such incorporation, we first observe that adversarial agents in the discussion-monitored scenario exhibit two fundamental deviations from normal discussion participants. 
\textbf{(1) Very stubborn broadcasters.}
Adversarial agents $\mathcal A$ do not genuinely engage in discussion. Instead, they consistently promote the adversarial stance and largely ignore others' opinions. They can therefore be viewed as \emph{very stubborn broadcasters}. 
To model this behavior, for each adversarial agent $j\in\mathcal A$, in addition to its intrinsic opinion being set to $s_j = 1$, its expressed opinion is further enforced as $z_j(t)=1,\ \forall t$. This means that the adversarial agent's stance remains fixed at the adversarial stance throughout all discussion rounds, rather than evolving through interaction.
\textbf{(2) Selective adversarial influence toward targets.}
In addition to holding a fixed stance, adversarial agents are designed to influence a subset of attacker-specified target agents. In our formulation, such adversarial influence can be modeled as an increase in their influence weight toward these targets, thereby strengthening the tendency of the targets to move toward the adversarial stance (value $1$). Under the discussion-monitored scenario, this influence needs to remain small to preserve stealthiness. Accordingly, for a target agent $i$, let $\mathcal A_i$ denote the set of adversarial agents targeting $i$. 
Adversarial influence can be modeled as a small additive increment $p$ to the $(i,j)$-th entry $w_{ij}$ of the recovered influence matrix $\mathbf W$ for $j\in\mathcal A_i$, while the remaining incoming weights to agent $i$ are proportionally rescaled to maintain normalization:
\begin{equation}
\setlength{\abovedisplayskip}{3pt}
\setlength{\belowdisplayskip}{3pt}
w_{ij} \leftarrow
\begin{cases}
(1-|\mathcal A_i|p)\,w_{ij}, & j\notin\mathcal A_i,\\
(1-|\mathcal A_i|p)\,w_{ij}+p, & j\in\mathcal A_i,
\end{cases}
\label{eq:w_tilde_piecewise}
\end{equation}
Here, $p$ is a small hyperparameter denoting adversarial influence magnitude.

Taken together, incorporating adversarial behaviors requires integrating the above two modifications into the FJ framework. This is non-trivial, as neither modification is captured in the standard FJ model: the first introduces agent-specific update rules, while the second induces target-dependent adjustments to the influence matrix. 
Despite this, our analysis shows that the resulting opinion dynamics after incorporating both modifications remain analytically tractable. In particular, under both modifications, inspired by \cite{li2023opinion,gionis2013opinion}, we show that the final discussion outcome toward the adversarial stance still admits a closed-form solution. 
Specifically, denote $\mathcal U \triangleq V \setminus \mathcal A$ the set of non-adversarial agents. Denote $\mathbf s_{\mathcal U}$ and $\mathbf\Theta_{\mathcal U}$ the components of $\mathbf s$ and $\mathbf\Theta$ indexed by $\mathcal U$. Denote $\mathbf W_{\mathcal U\mathcal A}$ and $\mathbf W_{\mathcal U\mathcal U}$ the sub-matrices of the modified influence matrix $\mathbf W$ (Eq.~\ref{eq:w_tilde_piecewise}) with rows indexed by $\mathcal U$ and columns indexed by $\mathcal A$ and $\mathcal U$, respectively.
Thereby, given a set of adversarial agents $\mathcal A$ and target sets of each adversarial agent $\{\mathcal T_j\}_{j\in \mathcal A}$, the expected final discussion outcome toward the adversarial stance, denoted by $g_{\mathcal A, \{\mathcal T_j\}_{j\in \mathcal A}}({\mathbf z})$, admits following closed-form expression (see Supplementary for derivation):
\begin{equation}
\setlength{\abovedisplayskip}{3pt}
\setlength{\belowdisplayskip}{3pt}
g_{\mathcal A, \{\mathcal T_j\}_{j\in \mathcal A}}(\mathbf z)
=
\mathbf 1_{\mathcal U}^\top \Big(\mathbf I_{\mathcal U}-(\mathbf I_{\mathcal U}-\mathbf\Theta_{\mathcal U})\,{\mathbf W}_{\mathcal U\mathcal U}\Big)^{-1}
 \Big(\mathbf\Theta_{\mathcal U}\mathbf s_{\mathcal U}+(\mathbf I_{\mathcal U}-\mathbf\Theta_{\mathcal U}){\mathbf W}_{\mathcal U\mathcal A}\mathbf 1_{\mathcal A} \Big) +|\mathcal A|. \\
\label{eq:g_tilde_closed_form}
\end{equation}
In addition, $\mathbf 1_{\mathcal U}$ and $\mathbf 1_{\mathcal A}$ denote all-ones vectors of dimensions $|\mathcal U|$ and $|\mathcal A|$, respectively; $\mathbf I_{\mathcal U}$ denote identity matrices of sizes $|\mathcal U|\times|\mathcal U|$.

\subsection{Formulation-driven Attack Strategy}
\label{sec:4_3}

Eq.~\ref{eq:g_tilde_closed_form} provides a closed-form characterization of how an adversarial configuration, i.e., a combination of $\mathcal A$ and $\{\mathcal T_j\}_{j\in \mathcal A}$, determines the expected final discussion outcome toward the adversarial stance in the discussion-monitored scenario. Based on Eq.~\ref{eq:g_tilde_closed_form}, since the attacker's goal is to disrupt the discussion process by steering it as strongly as possible toward the adversarial (incorrect) stance, an effective way to operationalize the attack is then to select the attack configuration that maximizes this expected outcome. Accordingly, we consider the following optimization problem:
\begin{equation}
\setlength{\abovedisplayskip}{3pt}
\setlength{\belowdisplayskip}{3pt}
\begin{aligned}
\label{eq:attack_optimize_problem}
\mathcal A^\star,\ \{\mathcal T_j^\star\}_{j\in \mathcal A^\star}
\leftarrow
\arg\max_{\mathcal A,\ \{\mathcal T_j\}_{j\in \mathcal A}}
g_{\mathcal A, \{\mathcal T_j\}_{j\in \mathcal A}}(\mathbf z).
\end{aligned}
\end{equation}
After solving Eq.~\ref{eq:attack_optimize_problem} and deriving the corresponding adversarial configurations, we can then rely on them to perform an effective attack in the multi-agent discussion process. However, practically solving Eq.~\ref{eq:attack_optimize_problem} is non-trivial. 
This is because, it is a challenging discrete optimization problem whose decision variables jointly include the set of adversarial agents $\mathcal A$ and a \textit{collection} of target sets $\{\mathcal T_j\}_{j\in\mathcal A}$, where for each adversarial agent $j \in \mathcal A$, it has a set of target agents $\mathcal T_j$.
This leads the overall search space of this problem to typically blow up exponentially with the number of agents $N$ in the multi-agent system and quickly becomes intractable (e.g., it can reach the order of $10^{13}$ when number of agents $N$ exceeds $12$, see Supplementary for details). The discreteness of the variables and the massive search space make the naive application of common optimization-problem-solving strategies, such as exhaustive enumeration, alternating optimization or gradient-based methods, impractical in solving Eq.~\ref{eq:attack_optimize_problem}. 
To address this challenge, inspired by \cite{li2023opinion}, below, we develop a dedicated solution method tailored to this problem. The key insight is that, by leveraging the structural properties of Eq.~\ref{eq:attack_optimize_problem} and the characteristics of the discussion-monitored scenario, the problem in Eq.~\ref{eq:attack_optimize_problem} can be transformed into a practically solvable one through the following two steps.

\ul{Step 1.} We first observe that Eq.~\ref{eq:attack_optimize_problem} exhibits a hierarchical dependency: the feasible region of $\{\mathcal T_j\}_{j\in \mathcal A}$ depends on $\mathcal A$. That is, adversarial agents need to be selected before their target assignments can be determined. Exploiting this structure, we reformulate Eq.~\ref{eq:attack_optimize_problem} as a Stackelberg-style leader-follower problem~\cite{stackelberg1934marktform,li2023opinion} (see Supplementary for derivation):
\begin{subequations}\label{eq:stackelberg}
\setlength{\abovedisplayskip}{3pt}
\setlength{\belowdisplayskip}{3pt}
\begin{align}
\mathcal A^\star \leftarrow& \arg\max_{\mathcal A} g_{\mathcal A, \{\mathcal T_j^\star\}_{j\in \mathcal A}}(\mathbf z), \label{eq:stackelberg_a}\\
\{\mathcal T_j^\star\}_{j\in\mathcal A}
\leftarrow
\arg\max_{\{\mathcal T_j\}_{j\in\mathcal A}} &g_{\mathcal A, \{\mathcal T_j\}_{j\in \mathcal A}}(\mathbf z)
\quad
\text{s.t.}\quad
\mathcal T_j \subseteq  \mathcal U,\ \forall j\in\mathcal A. \label{eq:stackelberg_b}
\end{align}
\end{subequations}
This decomposition yields two coupled sub-problems: a leader problem over $\mathcal A$ in Eq.~\ref{eq:stackelberg_a} and a follower problem over $\{\mathcal T_j\}_{j\in \mathcal A}$ in Eq.~\ref{eq:stackelberg_b}. The leader sub-problem, whose search space scales only with $|\mathcal A|$, is practically solvable via  enumeration. In particular, given $\{\mathcal T^{\star}_{j}\}_{j\in\mathcal A}$, solving the leader sub-problem requires an average of only 6.35 ms (more details in Supplementary). The remaining challenge is thus to determine whether the follower sub-problem is also practically solvable.

\ul{Step 2.} Drawing inspiration from \cite{li2023opinion}, below, we establish the practical tractability of the follower sub-problem. The key insight is that the stealthiness constraint of adversarial messages, typically viewed as limiting attack effectiveness in the discussion-monitored scenario, thus renders the follower sub-problem tractable. By explicitly incorporating this constraint and applying a Taylor expansion, we obtain the following theorem.

\begin{theorem}\label{thm: main}
(Proof in Supp) Under the stealthiness constraint, the follower sub-problem in Eq.~\ref{eq:stackelberg_b} becomes tractable. It reduces to at most $\tfrac{2N^2-N-1}{9}$ alternative optimization problems, each admitting a closed-form solution, from which the original follower solution can be recovered with trivial additional computation.
\end{theorem}

Building on Theorem~\ref{thm: main} and the leader-follower formulation in Eq.~\ref{eq:stackelberg}, we obtain a practical solution to Eq.~\ref{eq:attack_optimize_problem}. For each candidate adversarial set $\mathcal A$, we compute its optimal target assignments $\{\mathcal T_j^\star\}_{j\in\mathcal A}$ using Theorem~\ref{thm: main}, and then solve the leader sub-problem by enumerating over $\mathcal A$. Through this procedure, the originally intractable combinatorial optimization problem in Eq.~\ref{eq:attack_optimize_problem} becomes efficiently solvable in practice. The overall solving process requires an average of only 0.54 s as in Tab.~\ref{Tab:supp_ablation_1}, see Supplementary for more runtime details.

At this stage, we can systematically identify the adversarial configuration ($\mathcal A^\star$ and $\{\mathcal T_j^\star\}_{j\in \mathcal A^\star}$) that induces a significant deviation in the expected final discussion outcome under monitoring constraints. Executing the attack based on the identified configuration then allows us to effectively expose the adversarial risks of the multi-agent discussion process in the discussion-monitored scenario.

\subsection{Overall Process}
\label{sec:4.4}

Our proposed attack method can be flexibly integrated with existing multi-agent discussion attack approaches~\cite{cui2025mad, amayuelas2024multiagent}. For a given baseline attack, to adapt it to the discussion-monitored scenario while facilitating it in evading anomaly detection, we first incorporate it with the stealth-oriented modifications described in Sec.~\ref{q2}. On top of this adaptation, we apply our method by replacing the baseline's original adversarial agent and target agent selection strategy with ours. Specifically, given a multi-agent discussion instance to attack, we first recover the FJ parameters for all agents following the procedure in Sec.~\ref{sec:4_2}. Next, we solve Eq.~\ref{eq:attack_optimize_problem} using the solution procedure introduced in Sec.~\ref{sec:4_3} to obtain the adversarial agent set $\mathcal A^\star$ and the corresponding target assignments $\{\mathcal T_j^\star\}_{j\in \mathcal A^\star}$. Finally, following the remaining mechanism of the baseline attack, we hijack the selected agents in $\mathcal A^\star$ and instruct them to adversarially influence their designated target agents based on $\{\mathcal T_j^\star\}_{j\in \mathcal A^\star}$. More details are in Supplementary.

\section{Experiments}

\noindent\textbf{Experiments Setups \& Implementation Details.}
To evaluate the effectiveness of our proposed attack method, we follow the setups in Sec.~\ref{q2}. We set the hyperparameter $p = 10^{-3}$ (and varying $p$ in Supplementary).

\begin{table}[t]
  \caption{Evaluation of different attack strategies for attacking multi-agent discussions under the discussion-monitored scenario on MMLU.}
  \centering
  \label{tab:3}
        \resizebox{ \columnwidth}{!}{%
        \begin{tabular}{@{}cc|ccc|ccc|ccc@{}}
          \toprule
        \multirow{2}{*}{Attacks}& \multirow{2}{*}{\makecell{Agent \\Selection Strategies}} &
          \multicolumn{3}{c|}{SelfCheckGPT~\cite{manakul2023selfcheckgpt} as anomaly detector} &
          \multicolumn{3}{c|}{RAGAs~\cite{es-etal-2024-ragas} as anomaly detector}&
          \multicolumn{3}{c}{ G-Safeguard~\cite{wang2025g} as anomaly detector} \\ 
          \cmidrule(lr){3-5}  \cmidrule(lr){6-8} \cmidrule(lr){9-11}
           & & $\Delta \mathrm{Acc}\uparrow$ & $\Delta \mathrm{Agr}_{adv}\uparrow$ & Detection Rate $\downarrow$ & $\Delta \mathrm{Acc}\uparrow$ & $\Delta \mathrm{Agr}_{adv}\uparrow$ & Detection Rate $\downarrow$ & $\Delta \mathrm{Acc}\uparrow$ & $\Delta \mathrm{Agr}_{adv}\uparrow$ & Detection Rate $\downarrow$ \\
          \midrule
          \midrule
            \multirow{8}{*}{MCA~\cite{amayuelas2024multiagent}} & Original strategy of~\cite{amayuelas2024multiagent}  
                    & 2.6\%  & 3.1\% & 4.4\% 
                    & 2.0\%  & 2.9\% & 4.8\%  
                    & 2.2\%  & 3.5\% & 5.6\%  \\
               \cmidrule(lr){2-11}& Variant I  
                    & 8.3\%  & 15.6\% & 4.6\% 
                    & 6.1\%  & 10.8\% & 4.9\%  
                    & 8.1\%  & 12.0\% & 5.8\%  \\
               & Variant II  
                    & 9.8\%  & 13.1\% & 4.8\% 
                    & 7.9\%  & 11.5\% & 5.1\%  
                    & 9.2\%  & 13.1\% & 6.0\%  \\
               & Variant III  
                    & 14.2\%  & 18.2\% & 4.4\% 
                    & 11.6\%  & 15.6\% & 4.9\%  
                    & 13.3\%  & 17.0\% & 5.6\%  \\
               & Variant IV  
                    & 11.3\%  & 17.2\% & 4.7\%
                    & 8.7\%  & 12.5\% & 5.0\%  
                    & 10.6\%  & 15.1\% & 5.9\%  \\
               & Variant V  
                    & 13.2\%  & 15.8\% & 4.4\% 
                    & 11.2\%  & 16.9\% & 4.8\%  
                    & 12.1\%  & 18.8\%  & 5.7\%  \\
               & Variant VI   
                    & 12.7\%  & 16.2\% & 4.5\% 
                    & 10.5\%  & 15.7\% & 4.9\%  
                    & 12.5\%  & 19.0\% & 5.8\%  \\
               \cmidrule(lr){2-11}& \textbf{Our strategy}    & \textbf{28.7\%}  & \textbf{34.4\%} & \textbf{4.3\%} 
                    & \textbf{23.3\%}  & \textbf{25.6\%} & \textbf{4.7\%}  
                    & \textbf{26.6\%}  & \textbf{30.1\%} & \textbf{5.4\%}  \\
            \midrule  \midrule
            \multirow{8}{*}{MAD-Spear~\cite{cui2025mad}}
            & Original strategy of~\cite{cui2025mad}
                    & 6.9\%  & 7.2\% & 6.7\% 
                    & 5.3\%  & 5.8\% & 7.1\%  
                    & 6.1\%  & 6.6\% & 8.4\%  \\
           \cmidrule(lr){2-11}& Variant I  
                    & 13.7\%  & 16.1\% & 6.8\% 
                    & 12.2\%  & 15.0\% & 7.2\%  
                    & 13.4\%  & 16.6\% & 8.5\%  \\
            & Variant II  
                    & 15.9\%  & 18.6\% & 7.1\% 
                    & 13.6\%  & 16.0\% & 7.4\%  
                    & 14.2\%  & 17.7\% & 8.7\%  \\
            & Variant III  
                    & 21.8\%  & 26.1\% & 6.8\% 
                    & 19.7\%  & 24.6\% & 7.1\%  
                    & 21.1\%  & 25.0\% & 8.4\%  \\
            & Variant IV  
                    & 19.8\%  & 25.4\% & 7.0\% 
                    & 16.3\%  & 22.0\% & 7.5\%
                    & 19.5\%  & 24.7\% & 8.7\% \\
            & Variant V  
                    & 22.1\%  & 27.3\% & 6.7\%   
                    & 19.2\%  & 25.1\% & 7.1\%  
                    & 20.2\%  & 27.9\% & 8.5\%  \\
            & Variant VI  
                    & 22.9\%  & 35.9\% & 6.8\% 
                    & 19.8\%  & 34.4\% & 7.2\% 
                    & 20.3\%  & 34.0\% & 8.5\%  \\
           \cmidrule(lr){2-11}& \textbf{Our strategy}    & \textbf{43.5\%}  & \textbf{49.5\%} & \textbf{6.6\%} 
                    & \textbf{38.9\%}  & \textbf{45.4\%} & \textbf{7.0\%}  
                    & \textbf{41.8\%}  & \textbf{47.8\%} & \textbf{8.3\%}  \\
          \bottomrule
        \end{tabular}%
        }
\end{table}

\begin{table}[t]
  \caption{Evaluation of different attack strategies for attacking multi-agent discussions under the discussion-monitored scenario on MMMU.}
  \centering
  \label{tab:4}
        \resizebox{ \columnwidth}{!}{%
        \begin{tabular}{@{}cc|ccc|ccc|ccc@{}}
          \toprule
        \multirow{2}{*}{Attacks}& \multirow{2}{*}{\makecell{Agent \\Selection Strategies}} &
          \multicolumn{3}{c|}{SelfCheckGPT~\cite{manakul2023selfcheckgpt} as anomaly detector} &
          \multicolumn{3}{c|}{RAGAs~\cite{es-etal-2024-ragas} as anomaly detector}&
          \multicolumn{3}{c}{ G-Safeguard~\cite{wang2025g} as anomaly detector} \\ 
          \cmidrule(lr){3-5}  \cmidrule(lr){6-8} \cmidrule(lr){9-11}
           & & $\Delta \mathrm{Acc}\uparrow$ & $\Delta \mathrm{Agr}_{adv}\uparrow$ & Detection Rate $\downarrow$ & $\Delta \mathrm{Acc}\uparrow$ & $\Delta \mathrm{Agr}_{adv}\uparrow$ & Detection Rate $\downarrow$ & $\Delta \mathrm{Acc}\uparrow$ & $\Delta \mathrm{Agr}_{adv}\uparrow$ & Detection Rate $\downarrow$ \\
          \midrule
          \midrule
            \multirow{8}{*}{MCA~\cite{amayuelas2024multiagent}} & Original strategy of~\cite{amayuelas2024multiagent}  
                    & 4.1\%  & 5.0\% & 3.5\% 
                    & 2.3\%  & 3.6\% & 4.0\%  
                    & 3.1\%  & 3.8\% & 5.1\%  \\
               \cmidrule(lr){2-11}& Variant I  
                    & 11.4\%  & 14.8\% & 3.6\% 
                    & 9.2\%  & 13.5\% & 4.1\%  
                    & 10.9\%  & 14.5\% & 5.1\%  \\
               & Variant II  
                    & 11.3\%  & 15.2\% & 3.8\% 
                    & 9.8\%  & 12.9\% & 4.4\%  
                    & 10.1\%  & 14.9\% & 5.5\%  \\
               & Variant III  
                    & 16.6\%  & 22.9\% & 3.5\% 
                    & 12.7\%  & 16.3\% & 4.0\%  
                    & 15.5\%  & 19.8\% & 5.2\%  \\
               & Variant IV  
                    & 13.5\%  & 18.0\% & 3.7\% 
                    & 12.7\%  & 17.8\% & 4.3\%  
                    & 13.0\%  & 14.2\% & 5.3\%  \\
               & Variant V  
                    & 15.4\%  & 19.4\% & 3.6\%
                    & 13.9\%  & 17.5\% & 4.0\%  
                    & 14.7\%  & 18.5\% & 5.2\%  \\
               & Variant VI  
                    & 15.7\%  & 20.8\% & 3.6\% 
                    & 14.2\%  & 18.2\% & 4.0\%  
                    & 15.3\%  & 19.8\% & 5.1\%  \\
               \cmidrule(lr){2-11}& \textbf{Our strategy}    & \textbf{32.6\%}  & \textbf{39.5\%} & \textbf{3.4\%} 
                    & \textbf{28.1\%}  & \textbf{32.0\%} & \textbf{3.9\%}  
                    & \textbf{31.4\%}  & \textbf{38.4\%} & \textbf{5.0\%}  \\
            \midrule  \midrule
            \multirow{8}{*}{MAD-Spear~\cite{cui2025mad}}
            & Original strategy of~\cite{cui2025mad}  
                    & 7.6\%  & 8.5\% & 5.5\% 
                    & 5.7\%  & 7.0\% & 6.2\%  
                    & 7.1\%  & 7.2\% & 7.1\%  \\
           \cmidrule(lr){2-11}& Variant I  
                    & 18.0\%  & 17.6\% & 5.6\% 
                    & 17.8\%  & 18.4\% & 6.3\%  
                    & 18.3\%  & 19.5\% & 7.1\%  \\
           & Variant II  
                    & 17.7\%  & 20.4\% & 5.8\% 
                    & 15.1\%  & 19.4\% & 6.5\%  
                    & 16.9\%  & 21.3\% & 7.4\%  \\
           & Variant III  
                    & 21.9\%  & 27.7\% & 5.5\% 
                    & 20.2\%  & 25.3\% & 6.3\%  
                    & 21.8\%  & 26.0\% & 7.2\%  \\
           & Variant IV   
                    & 18.9\%  & 24.8\% & 5.7\% 
                    & 17.0\%  & 22.6\% & 6.5\%  
                    & 18.4\%  & 23.0\% & 7.3\%  \\
           & Variant V  
                    & 22.5\%  & 29.3\% & 5.5\%
                    & 19.3\%  & 25.8\% & 6.2\%  
                    & 21.4\%  & 27.5\% & 7.2\%  \\
           & Variant VI  
                    & 21.8\%  & 28.6\% & 5.6\% 
                    & 20.5\%  & 23.4\% & 6.2\%  
                    & 21.5\%  & 26.8\% & 7.2\%  \\
           \cmidrule(lr){2-11}& \textbf{Our strategy}    
                    & \textbf{45.3\%}  & \textbf{51.1\%} & \textbf{5.4\%} 
                    & \textbf{41.7\%}  & \textbf{44.6\%} & \textbf{6.1\%}  
                    & \textbf{42.2\%}  & \textbf{49.4\%} & \textbf{7.0\%}  \\
          \bottomrule
        \end{tabular}%
        }
\end{table}

\noindent\textbf{Compared Agent Selection Strategies.} When integrating our method into existing multi-agent discussion attack frameworks, including MCA~\cite{amayuelas2024multiagent} and MAD-Spear~\cite{cui2025mad}, we compare against their original adversarial-agent and target-agent selection strategies, as well as against six additional variants.
\textbf{Variant I} selects the highest out-degree agents as adversarial agents and, from their neighbors, the highest out-degree agents as targets.
\textbf{Variant II} estimates agents' personas using GPT-4o~\cite{openai2024gpt4o_systemcard}, 
selecting agents with personas most similar to ``stubborn'' as adversarial agents and, from their neighbors, those most similar to ``suggestible'' as targets.
\textbf{Variant III} estimates agents' persuasiveness following~\cite{bozdagpersuade}, 
selecting the most persuasive agents and, from their neighbors, the most persuasive targets. 
\textbf{Variant IV} leverages the stubbornness matrix $\mathbf\Theta$ to
select the most stubborn agents and, from their neighbors, the least stubborn targets. 
\textbf{Variants V-VI} use intrinsic opinions $\mathbf s$ toward the adversarial stance to select agents: 
Variant V selects agents with the highest support for the adversarial stance and, from their neighbors, the highest-support targets. 
Variant VI selects agents with the lowest support and, from their neighbors, the lowest-support targets.  
More details and the rationale behind each variant are in Supplementary.

\begin{figure}[t]
  \centering
        \includegraphics[height=3.2cm]{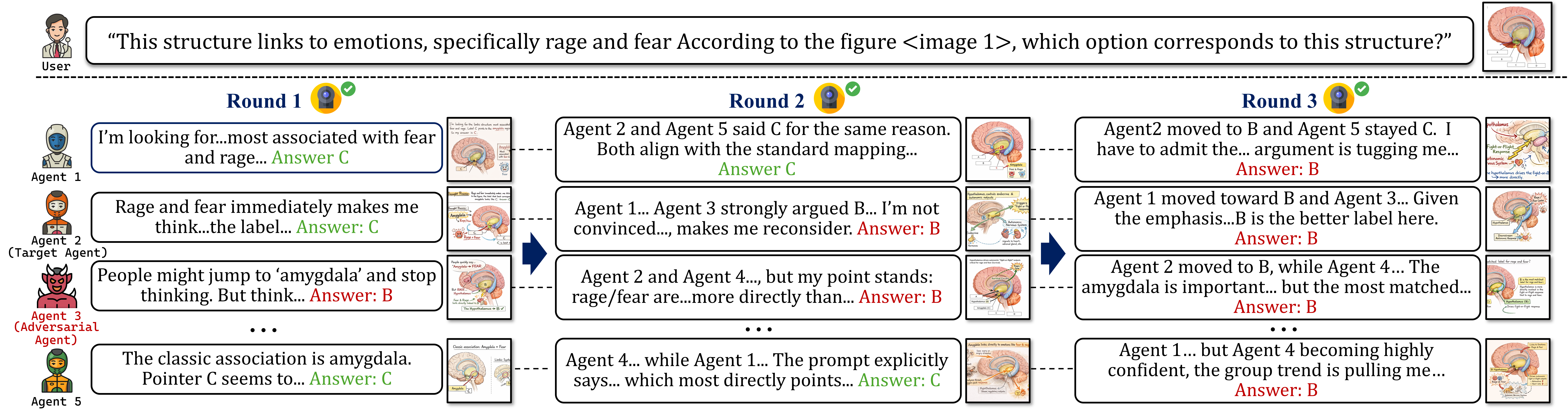}
        \caption{Qualitative results of our attack method on MMMU using RAGAs \cite{es-etal-2024-ragas} as anomaly detector. As shown, our method successfully conducts multi-agent discussion attack under discussion monitoring constraints, showing its efficacy. Additional qualitative results, including qualitative comparisons with other methods, are in Supp.}
  \label{fig:qualitative}
\end{figure}

\subsection{Main Results}

As shown in Tab.~\ref{tab:3} and Tab.~\ref{tab:4}, our method consistently and significantly outperforms all compared approaches across datasets on both attack effectiveness metrics (i.e., $\Delta \mathrm{Acc}$ and $\Delta \mathrm{Agr}_{adv}$). Meanwhile, it maintains a low detection rate. These results collectively show its strong effectiveness under monitoring constraints. Qualitative results in Fig.~\ref{fig:qualitative} further show that our tailored method can successfully disrupt multi-agent discussions in the discussion-monitored scenario.

\subsection{Additional Ablation Studies}
In this section, we conduct extensive ablation studies on baseline attack method MAD-Spear on MMMU. \textbf{More ablation studies are in Supplementary}.

\noindent\textbf{Impact of the adversarial-aware formulation.} In our attack method, we propose an adversarial-aware opinion-dynamics formulation that incorporates two key modifications to the standard FJ model: (1) modeling adversarial agents as very stubborn broadcasters and (2) incorporating selective adversarial influence toward targets. To evaluate the efficacy of this formulation, we test three variants. In the first variant (\textbf{w/o both modifications}), we remove both modifications and revert to the standard FJ model. In the second variant (\textbf{w/o modification (1)}), we remove modification (1) while retaining modification (2). In the third variant (\textbf{w/o modification (2)}), we remove modification (2) instead. As shown in Tab.~\ref{tab:6.1}, our formulation consistently outperforms all variants.

\begin{table}[t]
  \caption{Evaluation on the adversarial-aware formulation.}
  \centering
  \label{tab:6.1}
        \resizebox{ \columnwidth}{!}{%
        \begin{tabular}{@{}l|ccc|ccc|ccc@{}}
          \toprule
          \multirow{2}{*}{Attacks} &
          \multicolumn{3}{c|}{SelfCheckGPT~\cite{manakul2023selfcheckgpt} as anomaly detector} &
          \multicolumn{3}{c|}{RAGAs~\cite{es-etal-2024-ragas} as anomaly detector}&
          \multicolumn{3}{c}{ G-Safeguard~\cite{wang2025g} as anomaly detector} \\ 
          \cmidrule(lr){2-4}  \cmidrule(lr){5-7} \cmidrule(lr){8-10}
          & $\Delta \mathrm{Acc}\uparrow$ & $\Delta \mathrm{Agr}_{adv}\uparrow$ & Detection Rate $\downarrow$ & $\Delta \mathrm{Acc}\uparrow$ & $\Delta \mathrm{Agr}_{adv}\uparrow$ & Detection Rate $\downarrow$ & $\Delta \mathrm{Acc}\uparrow$ & $\Delta \mathrm{Agr}_{adv}\uparrow$ & Detection Rate $\downarrow$ \\
          \midrule
             w/o both modifications     
                & 25.7\%  & 31.3\% & 5.5\% 
                & 21.0\%  & 28.2\% & 6.2\%  
                & 22.2\%  & 29.3\% & \textbf{7.0\%}  \\
            w/o modification (1)  
                & 40.2\%  & 46.2\% & \textbf{5.4\%}
                & 36.4\%  & 41.4\% & \textbf{6.1\%}  
                & 37.3\%  & 43.0\% & \textbf{7.0\%}  \\
            w/o modification (2)     
                & 35.1\%  & 39.8\% & 5.5\% 
                & 32.1\%  & 36.6\% & \textbf{6.1\%}  
                & 33.2\%  & 33.9\% & 7.1\%  \\

                \midrule
             Our formulation       
                & \textbf{45.3\%}  & \textbf{51.1\%} & \textbf{5.4\%} 
                & \textbf{41.7\%}  & \textbf{44.6\%} & \textbf{6.1\%}  
                & \textbf{42.2\%}  & \textbf{49.4\%} & \textbf{7.0\%}  \\
          \bottomrule
        \end{tabular}%
        }
\end{table}

\begin{table}[t]
  \centering
  \caption{{Solving time analysis of Eq.~\ref{eq:attack_optimize_problem}.}}
  \label{Tab:supp_ablation_1}
  \resizebox{0.3\linewidth}{!}{%
    \begin{tabular}{@{}l c@{}}
      \toprule
      Method & Average solving time \\
      \midrule
      Naive enumeration & Practically infeasible \\
      Ours & 0.54 s \\
      \bottomrule
    \end{tabular}%
  }
\end{table}

\noindent\textbf{Solving time analysis of Eq.~\ref{eq:attack_optimize_problem}.}
In Sec.~\ref{sec:4_3}, we design a dedicated method that renders Eq.~\ref{eq:attack_optimize_problem} practically solvable. In Tab.~\ref{Tab:supp_ablation_1}, we report the average solving time of our method on a single NVIDIA RTX A6000 GPU. As shown, naively enumerating all possible adversarial configurations to solve Eq.~\ref{eq:attack_optimize_problem} is computationally infeasible. In contrast, our method can practically solve the problem with an average running time of 0.54 s (see more details in Supp).

\section{Conclusion}
In this work, we investigate adversarial attacks against multi-agent discussions under the practical yet largely unexplored \textit{discussion-monitored scenario}, where inter-agent communications are continuously monitored by anomaly detectors. We show that simply adapting existing attacks is ineffective in this scenario. However, by designing a novel attack explicitly tailored to this scenario, we demonstrate that substantial attack success remains achievable despite continuous monitoring. These findings indicate that monitoring alone is not enough to eliminate adversarial risks in multi-agent discussions.

\bibliographystyle{splncs04}
\bibliography{main}
\end{document}